\documentclass[12pt]{article}

\usepackage{graphicx,color}
\usepackage[superscript,sort,compress,noadjust]{cite}
\usepackage{amsmath,amssymb,latexsym}
\usepackage{authblk}
\newcommand\blfootnote[1]{%
  \begingroup
  \renewcommand\thefootnote{}\footnote{#1}%
  \addtocounter{footnote}{-1}%
  \endgroup
}

\title{The Conformation of Interfacially Adsorbed Ranaspumin-2 is an Arrested State on the Unfolding Pathway}
\date{}
\author[a, $\dagger$]{Ryan J. Morris}
\author[a, $\dagger$]{Giovanni B. Brandani}
\author[b]{Vibhuti Desai}
\author[b]{Brian O. Smith}
\author[a]{Marieke Schor}
\author[a,*]{Cait E. MacPhee}
\affil[a]{\footnotesize{School of Physics \& Astronomy, University of Edinburgh}}
\affil[b]{Institute of Molecular Cell and Systems Biology, University of Glasgow, Joseph Black Building, Glasgow G12 8QQ UK}
\affil[*]{To whom correspondence should be addressed: School of Physics \& Astronomy, University of Edinburgh, James Clerk Maxwell Building, Peter Guthrie Tait Road, Edinburgh, EH9 3FD, Tel: +44 (0) 131 651 7067, Fax: +44 (0) 131 650 5902, Email: cait.macphee@ed.ac.uk}
\affil[$\dagger$]{Contributed equally to this work.}

\begin{document}

\maketitle

\begin{abstract}

Ranaspumin-2 (Rsn-2) is a surfactant protein found in the foam nests of the t\'{u}ngara frog. Previous experimental work has led to a proposed model of adsorption which involves an unusual clam shell-like `unhinging' of the protein at an interface. Interestingly, there is no concomitant denaturation of the secondary structural elements of Rsn-2 with the large scale transformation of its tertiary structure. In this work we use both experiment and simulation to better understand the driving forces underpinning this unusual process. We develop a modified G\={o}-model approach where we have included explicit representation of the side-chains in order to realistically model the interaction between the secondary structure elements of the protein and the interface. Doing so allows for the study of the underlying energy landscape which governs the mechanism of Rsn-2 interfacial adsorption. Experimentally, we study targeted mutants of Rsn-2, using the Langmuir trough, pendant drop tensiometry and circular dichroism, to demonstrate that the clam-shell model is correct. We find that Rsn-2 adsorption is in fact a two-step process: the hydrophobic N-terminal tail recruits the protein to the interface after which Rsn-2 undergoes an unfolding transition which maintains its secondary structure. Intriguingly, our simulations show that the conformation Rsn-2 adopts at an interface is an arrested state along the denaturation pathway. More generally, our computational model should prove a useful, and computationally efficient, tool in studying the dynamics and energetics of protein-interface interactions.  

\end{abstract}

Keywords: molecular dynamics, coarse-grained simulations, interfacial proteins, protein folding, protein structure\\

\section{Introduction}

Protein-interface interactions are often non-specific and result in structural denaturation. While biologically undesirable, such non-specific interfacial activity has been exploited in many food technologies and processes, e.g. meringue formation. However, there are instances in which organisms have evolved to take advantage of proteins with specific interfacial activity to serve a specific biological purpose. Some examples of these are the anti-microbial lipopeptide surfactin\cite{Lu2007}, pulmonary surfactants\cite{Head2003}, the hydrophobins \cite{Wosten1994}, latherin \cite{McDonald2009}, and BslA, a hydrophobin-like protein present in biofilms of \emph{B. subtilis} \cite{Hobley2013}. In some cases (the hydrophobins and BslA) the protein-interface interactions are highly specific, and can result in ordered and crystalline films \cite{Szilvay2007,Bromley2015}.  

\blfootnote{Abbrevations used: Rsn-2, Ranaspumin-2; CD, circular dichroism; RIME, refractive index matched emulsion; NMR, nuclear magnetic resonance; DLS, dynamic light scattering.}

Here we study a protein component found in the foam nests of the t\'{u}ngara frog (\emph{Engystomops pustulosus}) known as Ranaspumin-2 (Rsn-2). Rsn-2 is known to be highly surface active without any associated lipid component \cite{Fleming2009} and has an amino acid sequence that is unique and distinct from hydrophobins or other known interfacially-active proteins. The surface activity of proteins such as the hydrophobins is immediately apparent from their structure, which is highly amphiphilic. The structure of Rsn-2, however, gives no indication of the mechanism of interfacial activity. Previous work on Rsn-2 has shown that there is an 8-10 \AA, hydrophobically rich, protein surface layer projecting from the air/water interface. Intriguingly, the secondary structure elements of Rsn-2 were found to be preserved when adsorbed to the interface. To explain these observations Mackenzie \emph{et al.} suggested a model where Rsn-2 undergoes a `clam-shell'-like unhinging at the interface: the hydrophobic faces of the $\alpha$-helix and $\beta$-sheet, which form the hydrophobic core of the protein, swing apart and become anchored at the interface while maintaining native secondary structure \cite{Mackenzie2009}.

In this work we confirm, using experiments on targeted mutant variants of Rsn-2, that the protein does undergo a large scale conformational change at an interface whilst maintaining its secondary structure elements consistent with the model proposed by Mackenzie \emph{et al.} With this established, a greater question is posed: from an energetic perspective, how does Rsn-2 undergo this unusual conformational change and avoid interfacial unfolding? Here we use simulations to explore the thermodynamic driving forces of this mechanism. Fully atomistic simulations aimed at capturing the complete dynamics of this process are too computationally expensive. Therefore, we develop a highly efficient, structure-based coarse-grained model of Rsn-2 using a similar strategy to that of Cieplak \emph{et al.} \cite{Cieplak2014}. Our G\={o}-model ~\cite{Taketomi1975, Go1981} differs from the latter example in that we represent the side-chains explicitly in order to realistically model the interaction between the secondary structure elements of the protein and the interface, and enable a comparison with experiments.
Our results show that Rsn-2 adsorption is a 2-step process: the flexible N-terminal tail of Rsn-2 allows the protein to first `capture' the interface, which is then followed by a large scale conformational re-orientation. Notably, this conformational transition is not accompanied by loss of the native secondary structure. These results correlate well with our experimental findings and the `clam-shell' model of adsorption. Importantly, we find that the structure observed at an interface is an intermediate, arrested conformational state on the unfolding pathway. Our results show evolution has finely tuned the energetics of Rsn-2 adsorption to the edge of stability. More generally, this computational model could prove useful as a tool to understand the energetics of adsorption and model the conformational dynamics of proteins at any interface.  

\section{Methods}

\subsection{Experimental methods}

\subsubsection*{Pendant Drop Tensiometry}
Pendant drop experiments were performed on a Kr\"{u}ss EasyDrop tensiometer. Rsn-2 was diluted in MilliQ water and immediately placed in a syringe with a needle diameter of 1.83 mm.  Images of the pendant drop are captured by a CCD camera and Kr\"{u}ss software fits the Young-Laplace equation to the drop shape to determine the interfacial tension. 

\subsubsection*{Langmuir Trough}

A KSV-Nima Langmuir trough was used in these experiments with a total interfacial area of 170 mm$^2$. Surface pressures were measured using the Wilhelmy Plate method. MilliQ water was used as the subphase and cleaned by aspiration. Experiments were carried out once the surface pressure remained at 0.2 mN m$^{-1}$ or less during a full compression. For each protein sample, 13 $\mu$g of material was applied to the interface. The equilibration time for the WT and the disulphide variant of Rsn-2 were 3 hours and 19 hours, respectively. 

\subsubsection*{Circular Dichroism}

Circular dichroism experiments were performed on a Jasco-810 spectropolarimeter. Refractive index matched emulsions (RIMEs) were made by first preparing a 20\% v/v decane emulsion with 0.2 mg ml$^{-1}$ Rsn-2. The emulsion was mixed for 1 minute using a rotor stator at 30,000 RPM. The emulsion was washed three times in order to remove any residual protein not adsorbed to an oil/water interface. This was done by allowing the emulsion to cream, a portion of supernatant was removed and replaced with buffer, then re-emulsified. Finally, supernatant was removed and replaced with glycerol such that the final wt\% of glycerol was 59\%. The emulsion was then re-mixed for 5 minutes using the rotor stator and then allowed to cream. The cream was then placed in a 1 mm mm path length quartz cuvette and the CD spectrum measured. 

\subsection{Coarse-grained computational model of Rsn-2}

In our model each amino-acid is represented by a backbone bead centred on the C$_{\alpha}$ atom and by a side-chain bead centred on the center-of-mass of the side-chain atoms of the residue (which is omitted in glycines).
The full Hamiltonian, which depends solely on the native conformation of the protein, is given by:
\begin{align}
\label{eq_hamiltonian}
V_{\rm protein} = \quad  & \sum_{\rm bonds} \epsilon_r (r_i-r_0)^2 
    +   \sum_{\rm angles} \epsilon_\theta (\theta_i-\theta_0)^2   \\ 
    + & \sum_{\rm dihedrals} \epsilon_\phi \lbrace \left[ 1-\cos(\phi_i-\phi_0) \right] + \frac{1}{2} \left[ 1-\cos(3(\phi_i-\phi_0)) \right] \rbrace \nonumber  \\
    + & \sum_{\rm contacts} \epsilon_{ij} \left[ \left(\frac{\sigma_{ij}}{r_{ij}}\right)^{12} - 2\left(\frac{\sigma_{ij}}{r_{ij}}\right)^6 \right] + \sum_{\rm non-contacts} \epsilon_{NC} \left(\frac{\sigma_{ij}}{r_{ij}}\right)^{12} \quad , \nonumber
\end{align}
where $\epsilon_r = 100\epsilon$, $\epsilon_\theta = 20\epsilon$, $\epsilon_\phi = 0.5\epsilon$, $\epsilon_{NC} = 0.01\epsilon$.
The equilibrium values of bond lengths $r_0$, angles $\theta_0$ and dihedral angles $\phi_0$ are computed from the first configuration of the Rsn-2 NMR ensemble~\cite{Mackenzie2009} (pdb id 2WGO, the parameters depend very weakly on the precise configuration used).
The attractive non-bonded interactions (contacts) are defined between each pair of coarse-grained beads for which any two atoms form a native contact in the NMR structure.
For instance, if residues $i$ and $j$ form a backbone hydrogen bond in the native structure, then we define an attractive interaction between the coarse-grained backbone beads of those two residues.
Similarly, a salt bridge or a hydrophobic contact between two side-chains will correspond to an attractive interaction between those two coarse-grained side-chain beads.
The native contacts are identified from the coordinates of the heavy atoms in the NMR structure using the shadow map method~\cite{Noel2010,Noel2012} with a cut-off radius of 6 \AA\ and shadowing radius of 1 \AA .
The strength of the contact energy $\epsilon_{ij}$ is set according to the following equation:
\begin{equation}
\epsilon_{ij} = \epsilon \frac{n_{ij}}{Z} \quad \text{where} \quad Z = n_{BB}+n_{BS}+n_{SB}+n_{SS} . 
\end{equation}
$n_{ij}$ is the number of native contacts between the atoms within the coarse-grained beads $i$ and $j$.
The sum on the right runs over all four possible interactions between the pair of residues to which the beads $i$ and $j$ belong: backbone-backbone, backbone-side-chain, side-chain-backbone and side-chain-side-chain.
The normalisation factor $Z$ makes sure that the total interaction energy between any pair of residues (with at least one native contact) is equal to $\epsilon$.
For instance, consider a pair of residues, $k$ and $l$; if the native structure contains two native-contacts between the backbone atoms of residue $k$ and the backbone atoms of residue $l$, and only one native contact between the backbone atoms of residue $k$ and the side-chain atoms of residue $l$, then the interaction energy between the two coarse-grained backbone beads will be $2/3 \epsilon$, whereas the energy between the backbone bead of residue $k$ and the side-chain bead of residue $l$ will be $1/3 \epsilon$.
This condition means that the interaction energies are distributed amongst the residues in the same way as in the standard structure-based model by Clementi et al.~\cite{Clementi2000}.
For the attractive non-bonded interactions, the equilibrium distance $\sigma_{ij}$ is set to the distance between beads $i$ and $j$ in the native structure; whereas for the repulsive non-bonded interactions (non-contacts), $\sigma_{ij}=0.5(\sigma_i+\sigma_j)$ (with $\sigma=4 \sqrt[3]{n_{\rm heavy}/4} $\AA , where $n_{\rm heavy}$ is the number of heavy atoms belonging to the coarse-grained bead). 

NMR spectroscopy could not resolve the N-terminal region (residues 1 to 16) of Rsn-2 due to its flexibility~\cite{Mackenzie2009}; therefore, the residues that are part of this region do not contribute to the native contacts, there is no potential acting on the dihedral angles, and the equilibrated structure deposited on the PDB database is only used for the calculation of the equilibrium bond lengths and angles.

With this choice of parameters, our coarse-grained model of Rsn-2 behaves like the structure-based model by Clementi et al.~\cite{Clementi2000}; in particular, the system is minimally frustrated and near the folding temperature T$_f$ the protein jumps between folded and unfolded states (see Fig.~\ref{fig_fes_step2}B).
We regard 0.9T$_f$ as ambient temperature; at this temperature the protein is folded in its native configuration and we never observe unfolding events (in bulk).

The interaction between the side-chain beads of the protein and an ideal water/oil interface perpendicular to $\hat{z}$ is modelled by the potential:
\begin{align}
\label{eq_interface}
V_{\rm interface} = & \quad \sum_{\rm hydrophobic} \epsilon_i \left[ \frac{1}{2}\left(\frac{\sigma_w }{r_{iz}}\right)^9 - \frac{3}{2}\left(\frac{\sigma_w }{r_{iz}}\right)^3 \right]  \\
                    & + \sum_{\rm hydrophilic} \epsilon_i \left(\frac{\sigma_w }{r_{iz}}\right)^9 \quad , \nonumber
\end{align}
where $\sigma_w =8$\AA\ and $r_{iz}$ is the distance along the $\hat{z}$ direction between side-chain $i$ and the interface. A side-chain is considered hydrophobic/hydrophilic if its cyclohexane-to-water free energy of transfer~\cite{Radzicka1988} is positive/negative; the interaction energy $\epsilon_i$ is proportional to the absolute value of this free energy, and the proportionality constant is set such that the sum of the attractive hydrophobic energies equals $0.65 \epsilon_{\rm folding}$, where $\epsilon_{\rm folding}$ is the sum of the contact energies $\epsilon_{ij}$ within the protein. This is the smallest attractive interaction that introduces a qualitative change in the folding landscape of the protein, and it provides a physical picture of the adsorption that is in agreement with our experimental observations (see results section). Note, the tuning of this interaction parameter is necessary for any G\={o}-model when there is an external component for the protein to interact with. For instance, this external component could be DNA \cite{Kenzaki2015} or a peptide \cite{Rogers2014}. In all cases there is experimental data which helps to guide the correct choice of interaction parameter. 
The backbone beads (excluding those of glycine, which are treated as hydrophobic side-chains) interact with the interface through a potential with the same form as that of hydrophobic side-chains, but with an energy equal to $\epsilon$ and a cut-off at $\sigma_w $, so that the force is only repulsive.

All simulations are performed in the NVT ensemble using the program\\
LAMMPS~\cite{Plimpton1995}.
The equations of motion are integrated using Langevin dynamics with a time-step $dt=0.001\sqrt{m/\epsilon}d_0$ and a relaxation time of $\tau_\text{rel}=2.6\sqrt{m/\epsilon}d_0$, where $m$ is the mass of each coarse-grained bead (all equal) and $d_0=3.8$ \AA\ is the typical bond length between consecutive backbone beads.
Periodic boundary conditions are applied along the $\hat{x}$ and $\hat{y}$ directions, whereas a repulsive wall perpendicular to $\hat{z}$ is placed 12 nm away from the interface.
In the simulations of spontaneous adsorption, the protein was initially placed 6 nm away from the interface and then left diffuse and adsorb at the interface until the system reaches a steady state.

Free energy profiles were computed from well-tempered metadynamics~\cite{Barducci2008} simulations.
Metadynamics is a sampling method that enhances the exploration of phase space by adding a history dependent potential bias to the Hamiltonian of the system. The bias depends on one or more collective variables $s$ and it is constructed in the following way: every time $\tau$, a Gaussian of width $\sigma$, height $w$ and centred at the current values of the collective variables is added to the bias potential; the height $w$ is equal to $w=w_0 e^{-V/k_B\Delta T}$, where $V$ is the bias potential evaluated at the current value of the collective variables, $w_0$ is the initial Gaussian height and $\Delta T$ is a parameter with the dimension of a temperature. At long times, the bias is related to the free energy of the system along the considered collective variables by $F(s)=-\frac{T+\Delta T}{\Delta T}V(s)$.

We will see in the result section that the adsorption of Rsn-2 at the interface occurs via a two-step mechanism: in the first step the protein adsorbs at the interface while maintaining its native fold, whereas in the second step it undergoes a partial unfolding that enables the protein to expose a larger number of hydrophobic residues to the interface.
To understand the roles of the different parts of the protein in each adsorption step, we considered three different protein mutants: the wild type (wt), the N-terminal deletion (d1-15) and a double disulphide bond mutant (2C-C).
The specific parameters used for these simulations depend on the considered system and protein mutant.
In our simulations we biased the dynamics of the system along the following collective variables:
\begin{itemize}
\item $d_{\rm interface}$, the distance between the center of mass of Rsn-2$_{16-96}$ (i.e. excluding the N-terminal tail) and the interface;
\item $d_{\rm tail}$, the distance along the $\hat{z}$ direction between the c.o.m. of Rsn-2$_{16-96}$ and the flexible N-terminal tail (residues 1-15 for wt Rsn-2 and 2C-C Rsn-2, and residue 16 for d1-15 Rsn-2)
\item $n_{\rm native}$, the number of native contacts within the protein, defined as:
\begin{equation}
n_{\rm native} = \sum_{\rm contacts} \epsilon_{ij}/\epsilon \frac{1-(r_{ij}/1.2\sigma_{ij})^8}{1-(r_{ij}/1.2\sigma_{ij})^{10}}
\end{equation}
$n_{\rm native}=224$ when the protein is folded in the native configuration, whereas it approaches zero in the unfolded state.
\end{itemize}
To study the free-energy landscape of the first step of the adsorption we biased $d_{\rm interface}$ and $d_{\rm tail}$; the use of both coordinates for the bias is necessary to avoid hysteresis in the reconstruction of the profile.
For this step alone we also defined a virtual harmonic wall that acts on the variable $n_{\rm native}$ and keeps it above a minimum value ($n_{\rm native,min}$=168) to prevent the unfolding of the protein; the settings are chosen to allow the natural fluctuations within the first metastable state of the adsorption.
In order to sample the second step of the adsorption we biased $d_{\rm interface}$ and $n_{\rm native}$, and we introduce a repulsive wall acting on the coordinates of all coarse-grained beads further that 6 nm from the interface, in order to prevent the complete desorption of the protein.
The free energy surfaces along the fraction of secondary structure and hydrophobic core contacts were obtained using the re-weighting method described by Bonomi \textit{et al.}\cite{Bonomi2009a}.
The free energy of the protein in bulk at the folding temperature was obtained using a bias on $n_{\rm native}$ only.
In all metadynamics simulations, we set $w_0=0.5k_BT$ and $\tau=500dt$.
The value of $\Delta T$ was varied between $5T$ and $19T$ depending on the system (high $\Delta T$ is needed to cross high free energy barriers, but it increases the error on the free energy if it is chosen too high): $\Delta T_\text{wt,step1}=\Delta T_\text{2C-C,step1}=\Delta T_\text{2C-C,step2}=19T$, $\Delta T_\text{wt,step2}=\Delta T_\text{d1-15,step1}=\Delta T_\text{d1-15,step2}=9T$, $\Delta T_\text{wt,bulk}=5T$.
The errors on the reported free energies are always within $\sim 1.5 k_B T$; convergence was assessed by looking at the difference in free energy between two relevant regions of the phase space (e.g. two local minima).
The parameter $\sigma$ is set equal to the standard deviation of the collective variable in the considered system:
$\sigma^{d_{\rm interface}}_\text{wt,step1}$=$\sigma^{d_{\rm interface}}_\text{d1-15,step1}$=$\sigma^{d_{\rm interface}}_\text{2C-C,step1}$=$0.1$\AA ,
$\sigma^{d_{\rm interface}}_\text{wt,step2}$=$\sigma^{d_{\rm interface}}_\text{d1-15,step2}$=$\sigma^{d_{\rm interface}}_\text{2C-C,step2}$=$0.05$\AA ,
$\sigma^{d_{\rm tail}}_\text{wt,step1}$=$\sigma^{d_{\rm tail}}_\text{2C-C,step1}$=$0.15$\AA ,
$\sigma^{d_{\rm tail}}_\text{d1-15,step1}$ $=0.2$\AA ,
$\sigma^{n_{\rm native}}_\text{wt,step2}$=$\sigma^{n_{\rm native}}_\text{d1-15,step2}$=$\sigma^{n_{\rm native}}_\text{2C-C,step2}$=$\sigma^{n_{\rm native}}_\text{wt,bulk}$=$6$.

\section{Results}

\subsubsection*{Initial Rsn-2 interfacial adsorption is diffusion limited}

\begin{figure}[htbp]
\begin{center}
\includegraphics[width=0.75\linewidth]{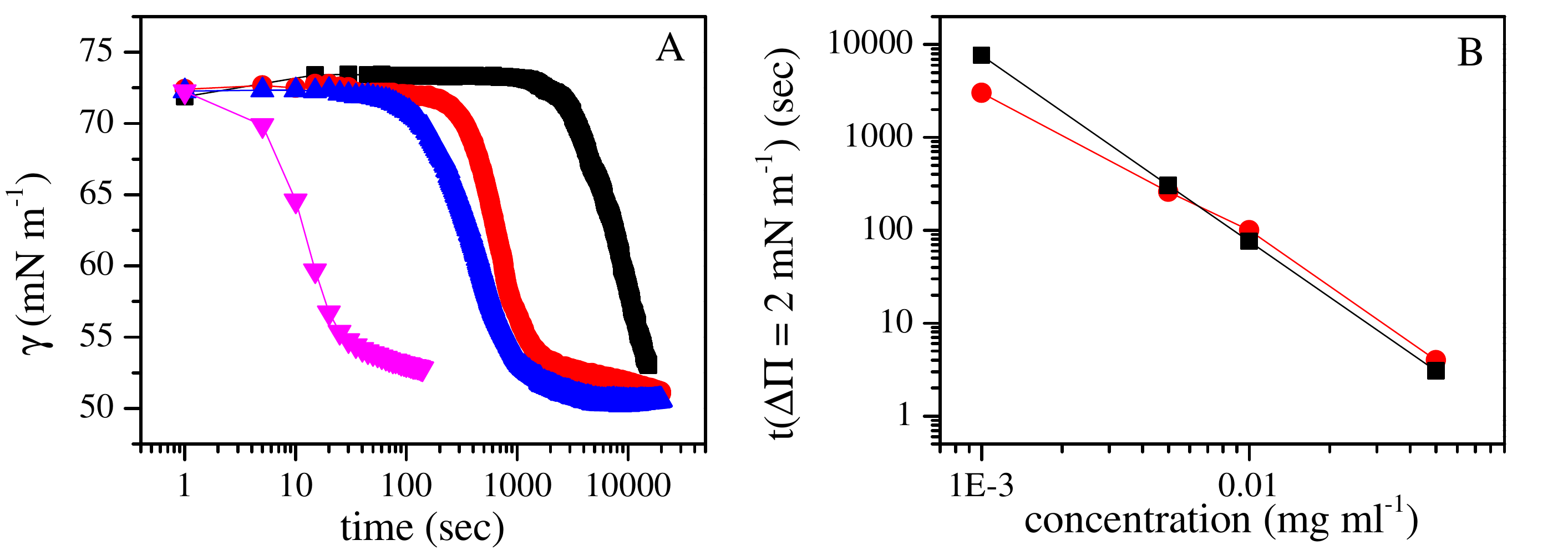}
\end{center}
\caption{Kinetics of Rsn-2 Adsorption to the Air-Water Interface. (A) Interfacial tension, $\Gamma$, was measured using pendant drop tensiometry. Figure shows $\Gamma$ as function of Rsn-2 concentration: 0.0012 mg ml$^{-1}$ (black), 0.005 mg ml$^{-1}$ (red), 0.01 mg ml$^{-1}$ (blue),0.05 mg ml$^{-1}$ (pink). (B) Black data points correspond to the theoretical time Rsn-2 would take to adsorb to an interface with a surface concentration of 0.95 mg m$^{2}$. This surface concentration corresponds to a $\Delta\Pi$ of 2 mN m$^{-1}$. The red data points are the times extracted from (A) for a change in surface pressure for the bare interface. The results show initial adsorption of Rsn-2 is well described by a diffusive model.  }
\label{fig:DiffLim}
\end{figure}

We utilized pendant drop tensiometry to study the kinetics of Rsn-2 surface activity at the air-water interface as a function of concentration. Upon expulsion from a needle tip, the shape of the Rsn-2 solution droplet is fit to the Young-Laplace equation which provides a measure of the surface tension. We define three kinetic regimes which characterize the absorption of protein to the interface: Regime I is when there is no apparent change in the interfacial tension, Regime II occurs when a sufficient amount of protein absorbs to the interface causing the interfacial tension to decrease, and Regime III is defined when the interfacial tension reaches a plateau. Regime III may show a further modest decrease in the surface tension which is often attributed to protein rearrangement \cite{Beverung1999}. Fig.~\ref{fig:DiffLim}A shows the results of these experiments. 

In order to discover whether Rsn-2 adsorption is purely diffusion limited or whether there exists some energy barrier to adsorption we calculated the time required for diffusion of the protein to the interface for comparison with experiment. The theoretical time it takes for a particle to diffuse and adsorb to an interface is

\begin{equation}
t=\frac{\pi}{D}\left(\frac{\Gamma(t)}{2c_b}\right)^2
\end{equation}

where $\Gamma(t)$ is surface concentration at time $t$, $c_{b}$ is the bulk concentration of protein, and $D$ is the diffusion coefficient \cite{Ward1946}. To parameterize this equation, $D$ and $\Gamma$ must be experimentally determined. First, we determine $D$ by utilizing the Stokes-Einstein relation $D=kT/6\pi R_{h}\eta$, where k is the Boltzmann constant, T is temperature and $\eta$ is the viscosity of the solution. We determined $R_h$ = 2.3 nm via dynamic light scattering (Fig. S1). Using this value we find $D$ = 9.3 x 10$^{-11}$ m$^2$s$^{-1}$. Second, to find $\Gamma(t)$ for a given $t$, we choose the time it takes for the interfacial tension to a reach a value $\gamma$ = 70.8 mN m$^{-1}$ ($\Delta\Pi$ =2 mN m$^{-1}$) from Fig. 1A. This value is chosen since it resides at the end of Regime I. We need to know the surface concentration of Rsn-2 when the interfacial tension is at this value. To do this, we use a Langmuir trough and measure surface-pressure isotherms (Fig.~S2A) to determine the area that corresponds to $\Delta\Pi$ = 2 mN m$^{-1}$. From this procedure we find $\Gamma$ = 0.95 mg m$^{-2}$. Using these values, we plot in Fig.~\ref{fig:DiffLim}B the theoretical time of diffusive adsorption against the experimental data from Fig.~\ref{fig:DiffLim}A. We find that initial Rsn-2 adsorption agrees well with a purely diffusive model and there is no energy barrier to adsorption (Fig.~\ref{fig:DiffLim}B). This is surprising given the proposed model in which Rsn-2 unhinges, since large conformational changes have previously been associated with subdiffusive rates of interfacial adsorption \cite{Bromley2015, Sengupta1999}. Later, we will explain the reason for this behaviour.

\begin{figure}[htbp]
\begin{center}
\includegraphics[width=0.75\linewidth]{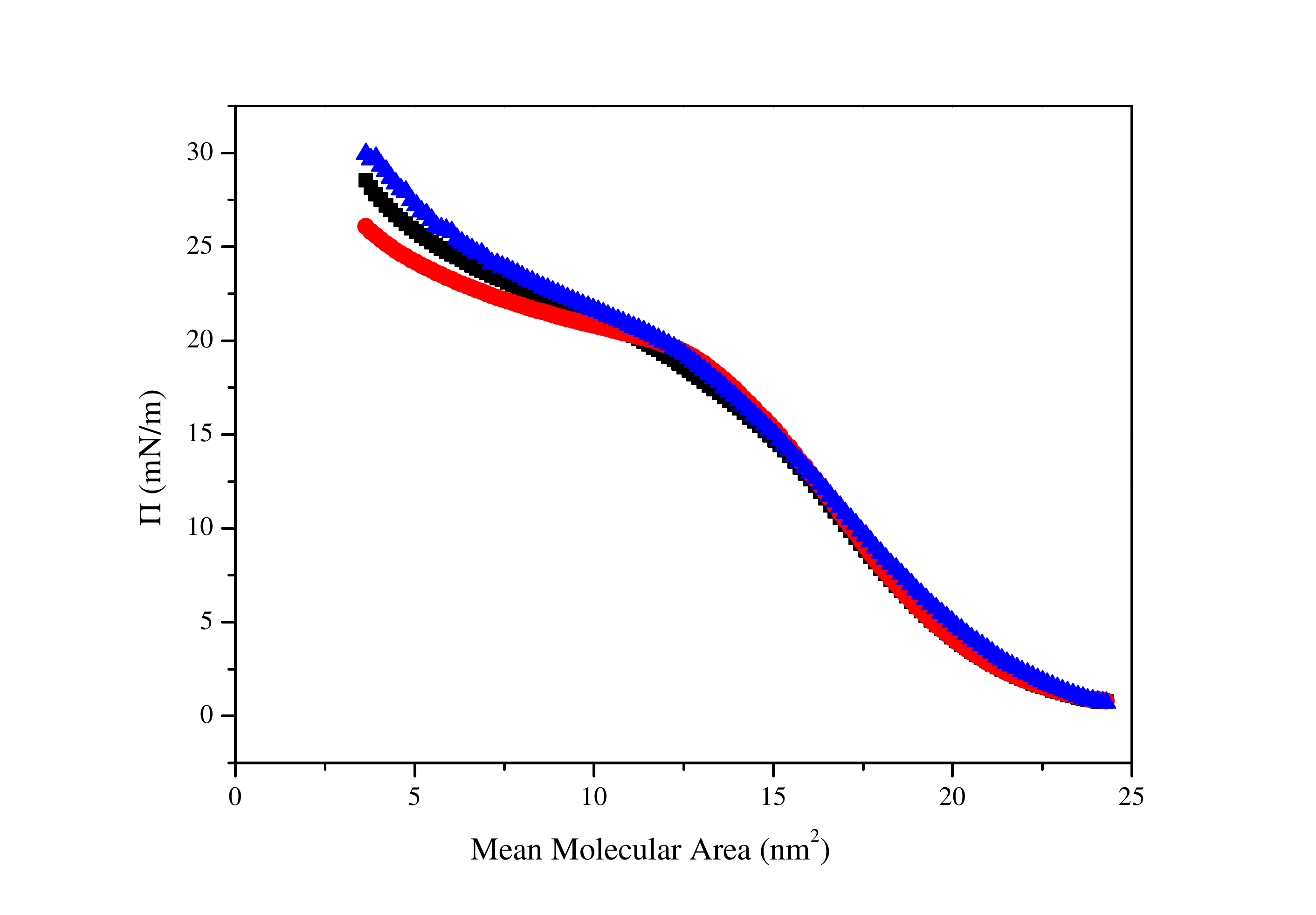}
\end{center}
\caption{Estimate of the size of Rsn-2 at an interface. Three independent area-pressure isotherms were obtained using a Langmuir trough. The turn over in the isotherm corresponds to surface saturation and was determined by finding the second derivative of the isotherm \cite{Aumaitre2011}. }
\label{fig:rsn2sizing}
\end{figure}

\subsubsection*{Estimating the Molecular Area of Rsn-2 at an Interface}

One can obtain a mean molecular area of a surface active molecule by measuring surface-pressure/area isotherms in a Langmuir trough.
A reliable value can be obtained when one can be assured that the total mass of the molecule that is applied at the surface remains at the surface.
For a water soluble protein such as Rsn-2 this condition cannot be met. However, we may estimate the mean molecular area of Rsn-2 after allowing for sufficient diffusion time for any protein lost to the subphase upon application to the interface.
Using the isotherms obtained from Fig.\ref{fig:rsn2sizing} and employing the observation that the inflection in the isotherm (minimum of the second derivative of the isotherm) corresponds to surface saturation (see SI for details), we obtain a mean molecular area $A_\text{mol}\approx$ 14 nm$^{2}$.

This result is in principle consistent with a crude estimate obtained from the hydrodynamic radius of the protein, which would give an area per molecule equal to $\pi R_h^2$=16.6 nm. However, this estimate does n ot take into account the fact that the protein is not a perfect hard sphere, and therefore, even in the absence of unfolding upon adsorption, the molecular area will depend on the precise orientation and conformation that the protein adopts at the interface.

One further observation is worth noting from these Langmuir trough experiments. Compression and expansion cycles indicated that Rsn-2 could be removed from the interface. Moreover, the protein lost from the interface through compression cycles could then diffuse back to the interface and bind again. This result indicates that Rsn-2 adsorption is a reversible process. For more details see the SI.   

\subsubsection*{Surface Activity and Structure of Rsn-2 Variants}

With these insights we produced two variants of Rsn-2. First, in order to probe the idea that the N-terminal region is responsible for `capturing' the interface a large portion of the N-terminus (residues 1-15) was deleted. Second, we wished to prevent the proposed unhinging of the protein by introducing two disulphide bridges between residues 19 and 46 and between residues 32 and 81 (henceforth referred to as 2C-C). The introduced disulphide bonds anchor the N- and C- terminal ends of the $\alpha$-helix to adjacent residues in the first and last strands of the $\beta$-sheet. It was found that the CD structure of these proteins in solution was identical to wild-type Rsn-2(data not shown). 

\begin{figure}[htbp]
\begin{center}
\includegraphics[width=0.75\linewidth]{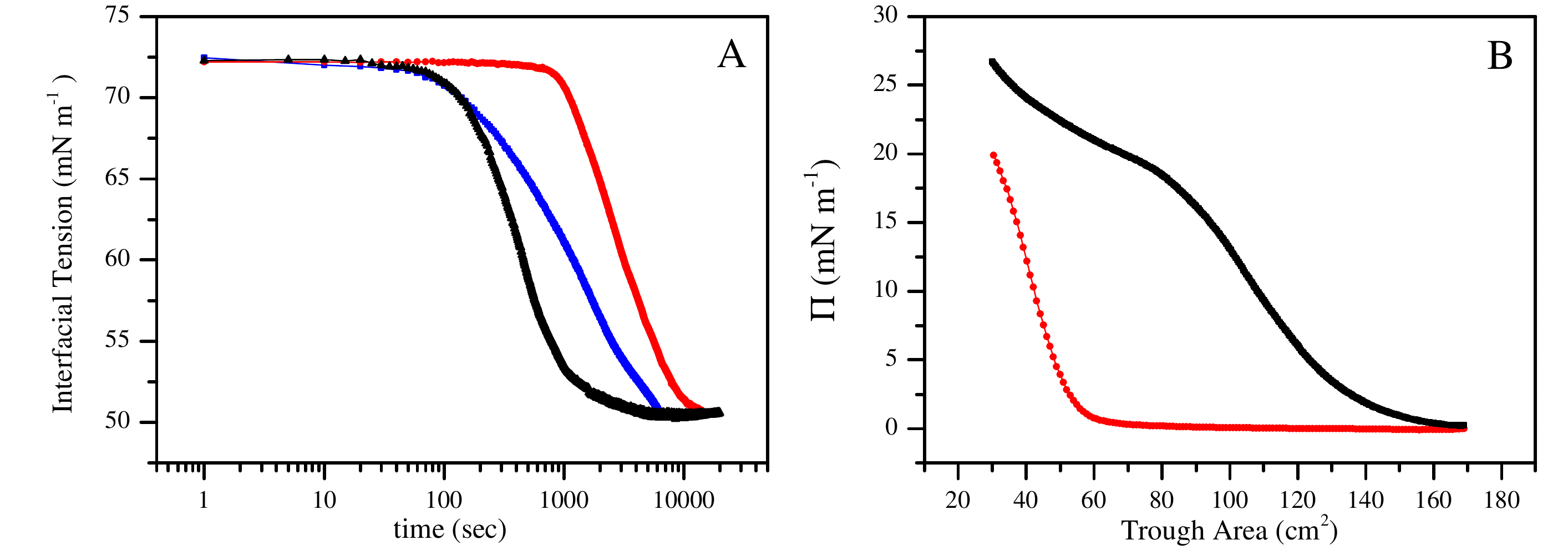}
\end{center}
\caption{Behavior of Rsn-2 variants at the air-water interface. (A) Pendant drop tensiometry of wild-type Rsn-2 (black), N-terminal deletion (red), and 2C-C mutant (blue). All concentrations were 0.01 mg ml$^{-1}$. (B) Surface pressure/area isotherms for wild-type Rsn-2 (black) and the 2C-C mutant (red). This result demonstrates that the 2C-C mutant occupies a significantly smaller area at the interface compared to wild-type.}
\label{fig:VariantFig}
\end{figure}

To confirm that the 2C-C mutant prevents the unhinging of Rsn-2 at the interface, we measured surface pressure/area isotherms in the Langmuir trough. We applied equivalent amounts of wild-type and mutant to the surface and allowed the system to equilibrate for 19 hours. We found that the area occupied by the 2C-C mutant was almost half as large as for wild-type Rsn-2 (Fig.\ref{fig:VariantFig}B). Finding an inflection point of the isotherm, as was done for wild-type, yields a mean molecular area of $\sim$ 7 nm$^2$, providing confirmation that the 2C-C mutant does restrict the unhinging of the protein at the interface.

We performed pendant drop tensiometry on these Rsn-2 variants and found that the Regime I time for the N-terminal deletion is approximately 10 times longer compared to wild-type. In contrast, the Regime I time for the 2C-C mutant is the same as wild-type Rsn-2 (Fig.\ref{fig:VariantFig}A). This result demonstrates the importance of the N-terminal tail in getting Rsn-2 to the interface. Note that the final interfacial tension is the same for the wild-type protein and two variants but the rate of interfacial tension change during Regime II for the 2C-C mutant is slower compared to wild-type and the N-terminal deletion. This is because the 2C-C mutant can not unhinge at the interface and only adsorbs via diffusion. It therefore requires more protein mass per unit area to achieve the equivalent surface tension change relative to wild type Rsn-2. 

If the same value of interfacial tension is reached for both the 2C-C variant and wild-type, why do we observe different isotherms for the 2C-C mutant and wild-type? The difference between the Langmuir trough and pendant drop experiments resides in the fact that for the pendant drop experiment, there is a large excess of protein in the bulk which eventually diffuses to the surface and lowers the IFT. However, in the trough case, there is no excess in the bulk. A limited mass of protein is applied such that all protein should only be present at the interface. Therefore, the 2C-C mutant, if it is not `un-hinging', should display the isotherm we observe, i.e. an isotherm shifted to smaller trough area compared to the wild-type.

\begin{figure}[htbp]
\begin{center}
\includegraphics[width=0.75\linewidth]{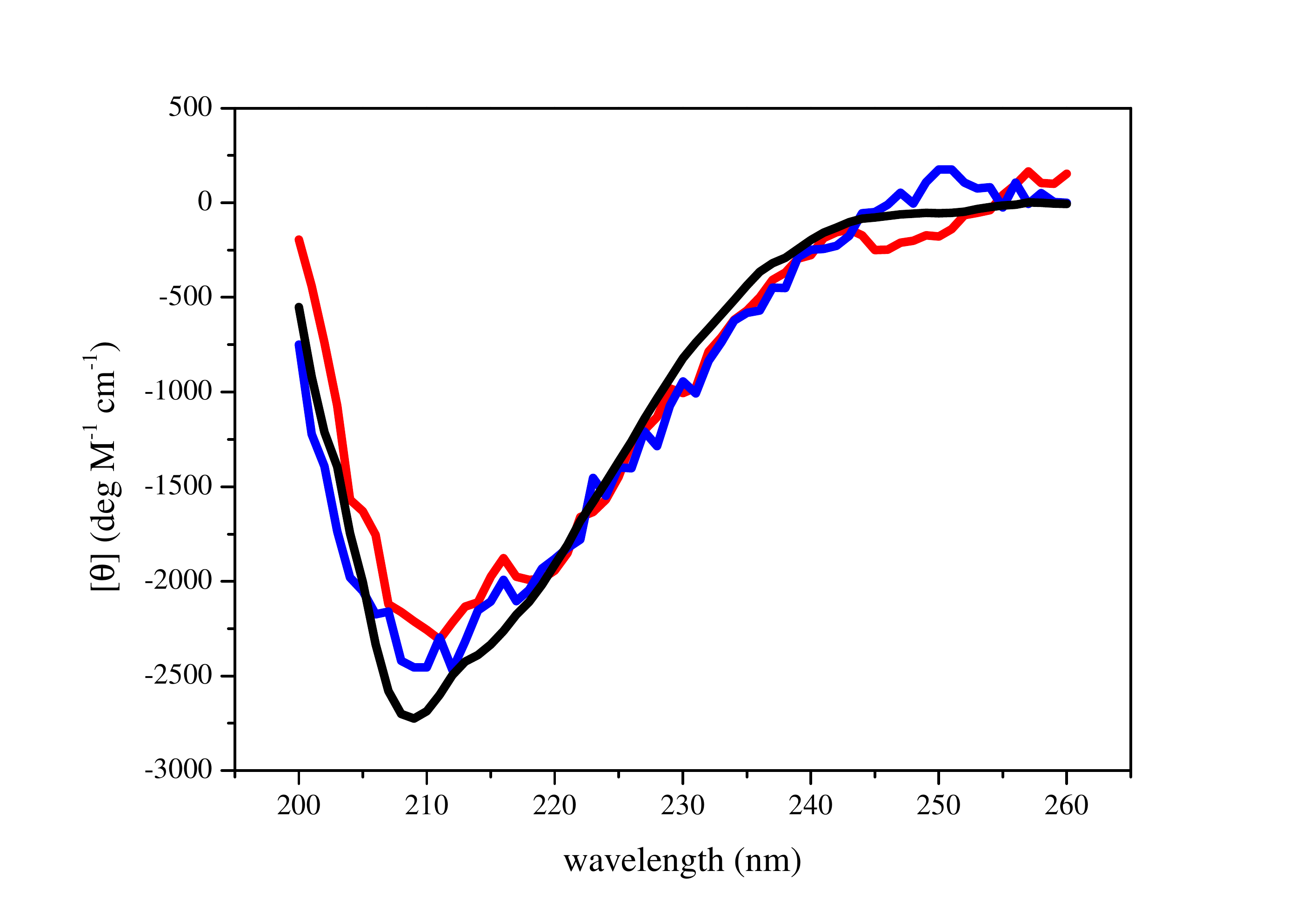}
\end{center}
\caption{Structure of Rsn-2 at the interface. The spectrum of wild-type Rsn-2 in solution (black), RIME of wild-type Rsn-2(red), and RIME of 2C-C mutant (blue). The spectra are normalized assuming all protein is present in the emulsion.   }
\label{fig:RIMECD}
\end{figure}

To study the structure of Rsn-2 at an interface, refractive index matched emulsions \cite{Husband2001} (RIMEs, see Materials \& Methods) of 0.2 mg ml$^{-1}$ wild-type Rsn-2 and the 2C-C variant were measured by CD. We could not perform this experiment with the N-terminal deletion variant due to poor emulsification, a failure that emphasizes the importance of the N-terminal for interfacial adsorption and activity. The result of this experiment is shown in Fig.\ref{fig:RIMECD}. For comparison, the CD spectrum of Rsn-2 in solution is also shown. We find that the RIME spectra for wild-type Rsn-2 and the 2C-C mutant are very similar that observed for the protein in aqueous solution. Based on this experiment, and previous work demonstrating a similar result \cite{Mackenzie2009}, we conclude that Rsn-2 retains its gross secondary structure elements upon adsorption to an interface.

\subsubsection*{Simulations Demonstrate Rsn-2 Interfacial Adsorption Is A 2-Step Process}

\begin{figure}[htbp]
\begin{center}
\includegraphics[width=0.5\linewidth]{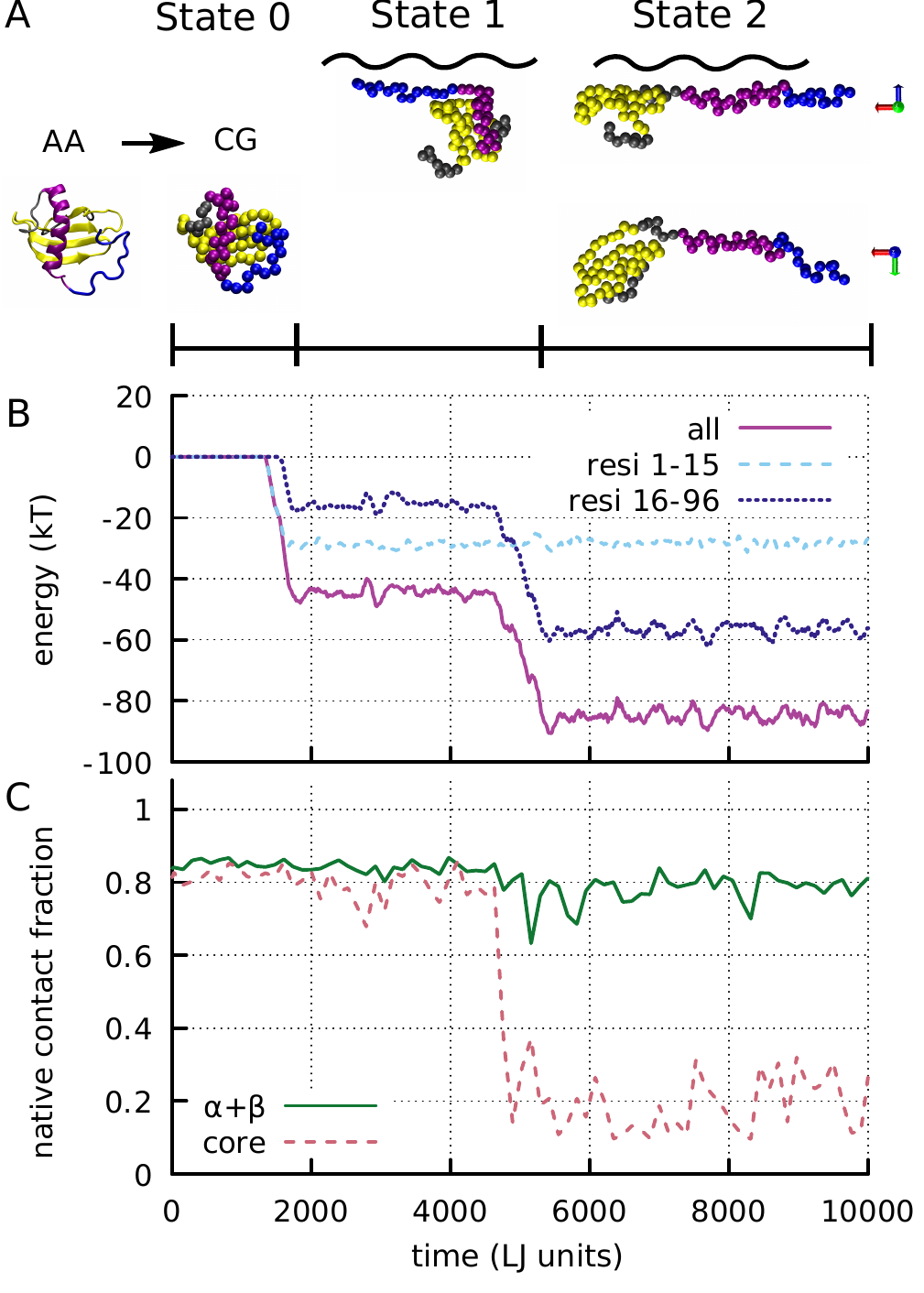}
\end{center}
\caption{
Energetics of adsorption of Rsn-2 to an ideal interface. A) Schematics of the coarse-graining (CG) from the all-atom NMR structure (AA), and typical Rsn-2 configuration at different adsorption stages: in bulk (state 0, corresponding to the native structure); with the flexible N-terminal region adsorbed at the interface (state 1); fully adsorbed and partially unfolded in side-view (top) and top-view (bottom) (state 2).
The N-terminal tail (residue ids 1-15) is shown in blue, the $\alpha$-helix (16-38) in purple, the $\beta$-sheet (45-88) in yellow, and the loop (39-44) and the C-terminal tail (89-96) in grey. The interface is represented by the black wavy line. 
For the CG structures we only show the $C_\alpha$ beads and not the side chains.
B) Energy of interaction with the interface for a typical adsorption event, showing the energetic contributions from the N-terminus (residues 1-15; dashed cyan line) and the globular region of the protein (residues 16-96; dashed blue line) to Rsn-2 adsorption (purple line).
C) Fraction of native contacts corresponding to the $\alpha$-helix and the $\beta$-sheet (green line) and those involved in the formation the hydrophobic core (dashed pink line).
The first adsorption step involves no loss of native contacts, whereas the second step involves the disruption of the hydrophobic core while the secondary structure elements are conserved.
}
\label{fig_adsorption_wt}
\end{figure}

Interfacial adsorption of proteins is, except in rare cases, associated with denaturation of both secondary and tertiary structures. We have found Rsn-2 undergoes a large scale tertiary conformational reorganization at an interface, while preserving its secondary structure. What is the energetic landscape that governs this unusual process? 

To answer this question we turn to molecular dynamics simulations. We developed a structure-based coarse-grained model to investigate the molecular mechanisms and energetics associated with Rsn-2 interfacial adsorption. We employ a modified G\=o-type model~\cite{Taketomi1975,Go1981} based on the NMR structure (PDB id: 2WGO) where we have coarse-grained the protein to include not only the C$_{\alpha}$ atom but also a side-chain bead. The Hamiltonian includes terms to account for the hydrophobicity or hydrophilicity of each coarse-grained side chain. This enabled us to study the competition between internal hydrophobic contacts within the core of the protein and external hydrophobic contacts between the protein and the interface.

Our molecular dynamics simulations show that the adsorption of Rsn-2 at an ideal water/oil interface proceeds via a two-step mechanism (Fig.~\ref{fig_adsorption_wt}): firstly the flexible N-terminal tail adsorbs at the interface while the rest of the protein maintains its native fold (state 1), and then the hydrophobic core of the protein unfolds, so that the two secondary structure elements (one $\alpha$-helix and one $\beta$-sheet) expose their hydrophobic side chains to the interface (state 2).
In the first state most of the interaction energy with the interface comes from the first 15 residues of the protein (Fig.~\ref{fig_adsorption_wt}B). This region has a very high hydrophobic content (6 Leucines, 1 Isoleucine and 1 Proline) and its flexibility enables the protein to adsorb without any barrier. These results explain the observation that Rsn-2 adsorption is diffusion limited (Fig. 1).

On the other hand, the energy of the second state is dominated by the interaction between the interface and side-chains in the globular part of the protein; reaching this state involves breakage of interactions in the native hydrophobic core (Fig.~\ref{fig_adsorption_wt}C) and therefore crossing of a free energy barrier.
It is interesting to note that the second step of the adsorption is not associated with the loss of secondary structure elements (Fig.~\ref{fig_adsorption_wt}C).
Importantly, both the $\alpha$-helix and the $\beta$-sheet have a well-defined hydrophobic dipole~\cite{Eisenberg1982}, and their reorientation at the interface is sufficient to optimise the interactions without any loss of secondary structure.

From our simulations, we also estimated the molecular area of Rsn-2 at the interface as the area of the projection on the $xy$-plane of the coarse-grained beads of the protein, each of them having a diameter of 4 \AA .
This analysis shows that the partial unfolding is accompanied by an increase in the molecular area, which is $A_\text{mol,state1}\simeq$9.9 nm$^2$ in the folded conformation and $A_\text{mol,state2}\simeq$13.3 nm$^2$ in the unhinged state. This observation helps to understand the difference between the area occupied by wild-type Rsn-2, expected to be unfolded at the interface, and the 2C-C mutant, expected to be folded.
$A_\text{mol,state2}$ is in quantitative agreement with the experimental molecular area for the wild type protein, whereas $A_\text{mol,state1}$ is slightly higher than the value measured for the mutant. We suggest that this difference might be due to the fact that in the experiments not all 2C-C mutants adsorb from the solution to the interface, due to the slower adsorption kinetics (Fig. 3A) and/or lower free energy of adsorption (see later). In other words, for calculating the area occupied, our assumption that 100\% of the 2C-C mutant has adsorbed to the interface is not correct.

\subsubsection*{Simulation of Rsn-2 Variants}

\begin{figure}[htbp]
\begin{center}
\includegraphics[width=0.5\linewidth]{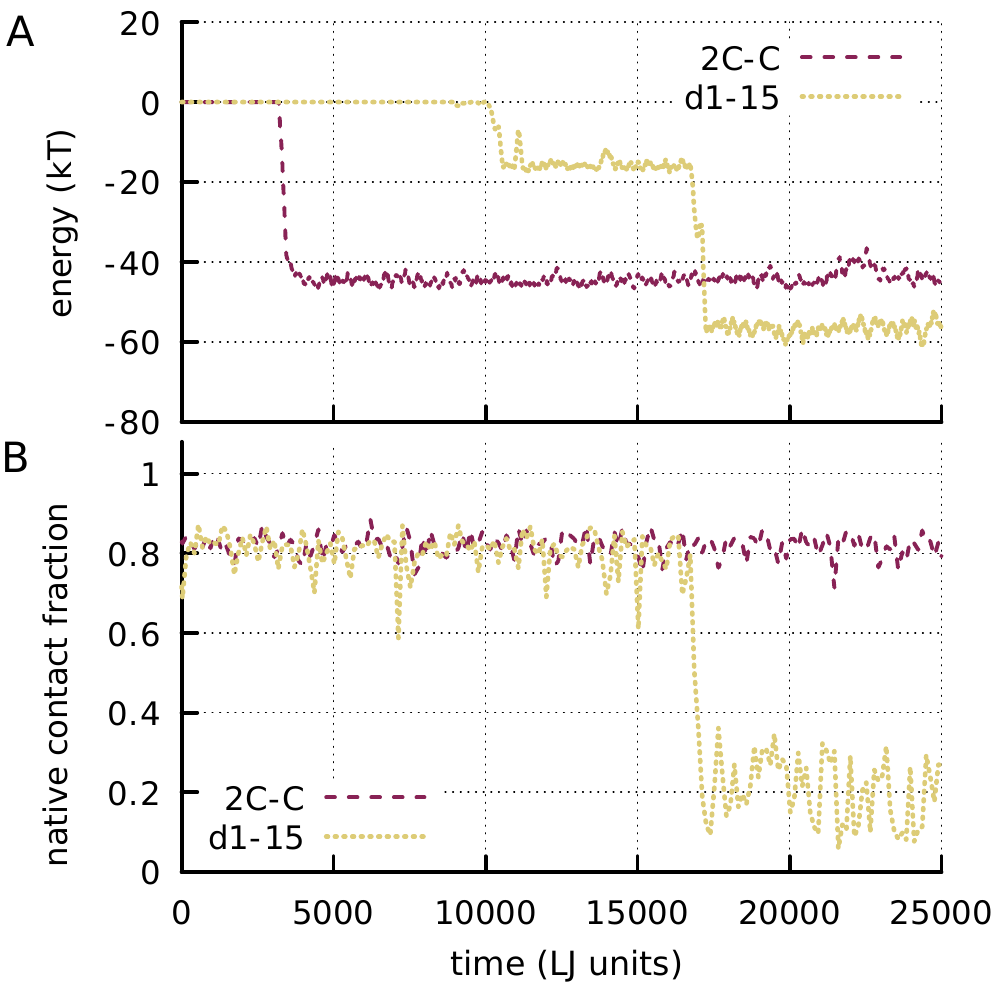}
\end{center}
\caption{
Energy of interaction with the interface and fraction of hydrophobic core native contacts for typical adsorption events observed for the d1-15 (yellow dotted lines) and 2C-C mutant (brown dashed lines). The deletion of the flexible tail reduces the interaction energy of the first adsorption state. The addition of the two disulphide bridges does not affect the first step but abolishes the unfolding of the hydrophobic core resulting in a significant reduction of the interaction energy.
}
\label{fig_adsorption_mut}
\end{figure}

Next, we perform simulations on the same two mutants studied in the experiments: in the first one we delete the first 15 residues of the N-terminal tail (d1-15), and in the second we apply two disulphide bridges between residues 19 and 46 and between residues 32 and 81.
Despite the lack of many native hydrophobic residues that were involved in the interaction with the interface (see Fig.~\ref{fig_adsorption_wt}B), the d1-15 mutant is still able to adsorb; this is due to the presence of other exposed hydrophobic residues, namely: V16, I17, L20, F21, V78 and V79.
Visual inspection of the simulations shows that the orientation of the mutant in the first step of the adsorption is the same as that for wild type Rsn-2.
However, Fig.~\ref{fig_adsorption_mut}A shows that the d1-15 deletion greatly reduces the interaction energy with the interface during the first adsorption state: this is only $\Delta E_{\text{d1-15,state1}}\simeq-16 k_B T$ for the mutant, whereas it is $\Delta E_{\text{wt,state1}}\simeq-44 k_B T$ for the wild type.
The deletion of the tail does not affect the second step of the adsorption, corresponding to the partial unfolding of the protein (Fig.~\ref{fig_adsorption_mut}B).
Conversely, the introduction of the two extra disulphide bonds does not affect the energetics of the first step of the adsorption (Fig.~\ref{fig_adsorption_mut}A), but it  prevents the unfolding of the hydrophobic core (Fig.~\ref{fig_adsorption_mut}B), which explains why this mutant occupies a smaller area at the interface than WT-Rsn-2 (Fig 3B).
For this mutant the adsorption is essentially arrested to the first state, where the protein interacts with the interface through the flexible N-terminal tail, whereas the other hydrophobic residues are mostly buried in the protein core.

\subsubsection*{Mapping the Free-energy Landscape of Interfacial Adsorption}

\begin{figure}[htbp]
\begin{center}
\includegraphics[width=0.6\linewidth]{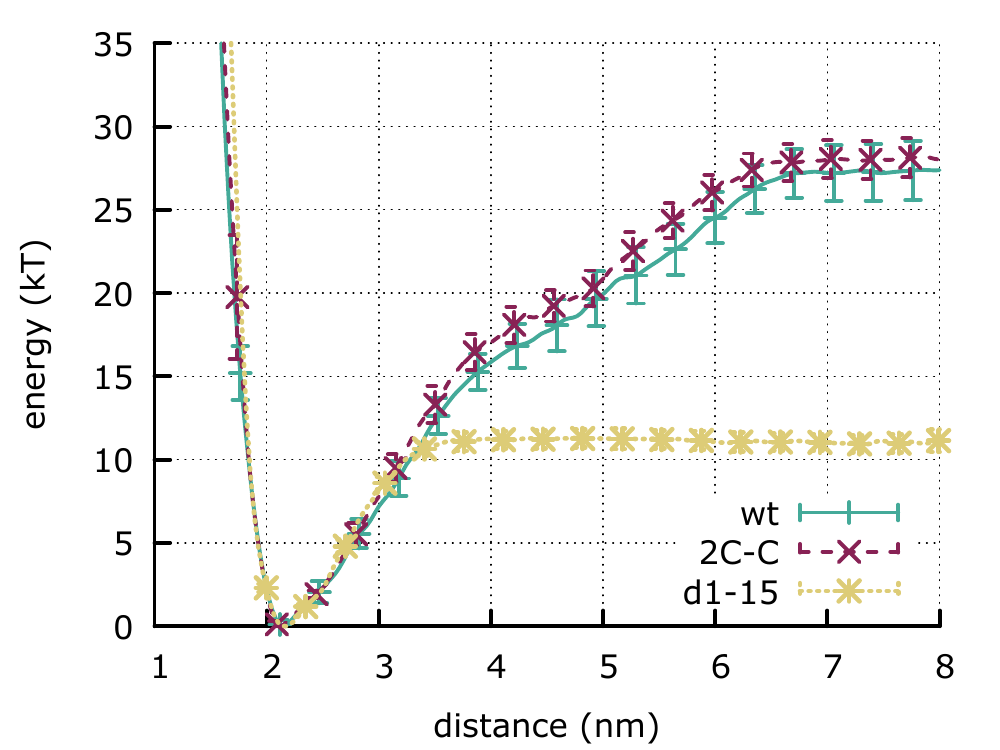}
\end{center}
\caption{Free energy of the first step of adsorption for wt (green), d1-15 (yellow) and 2C-C (brown) Rsn-2 as a function of the distance between the interface and the center of mass of the section of the protein common to all mutants (residues 16 to 96). The disulphide bridges have no effect on the profile, whereas the tail deletion reduces the free energy of adsorption to about a half.}
\label{fig_fes_step1}
\end{figure}

To better understand the energetics of this adsorption mechanism we employ the well-tempered metadynamics method~\cite{Barducci2008} to calculate the free energy landscape of the first and second steps, separately.
Figure~\ref{fig_fes_step1} displays the free energy profile of the system for the first step.
We calculated the free energy as a function of the distance between the protein and the interface.
The globular core of the protein is kept folded in the simulations to allow us to probe just this adsorption step.
The comparison between the mutants shows that the flexible N-terminal tail is responsible for most of the energy difference between the first adsorption state and the bulk state, and that the deletion of the tail reduces the range of the attraction from 4.5 nm down to 1.5 nm.
This range is defined as the distance between the free energy minimum of state 1 and the plateau of the bulk state (Fig.~\ref{fig_fes_step1}). This is consistent with the result found in Fig.~\ref{fig:VariantFig}A where the N-terminal deletion resulted in much longer initial adsorption times, while the 2C-C mutant is identical to wild-type

\begin{figure}[htbp]
\begin{center}
\includegraphics[width=0.9\linewidth]{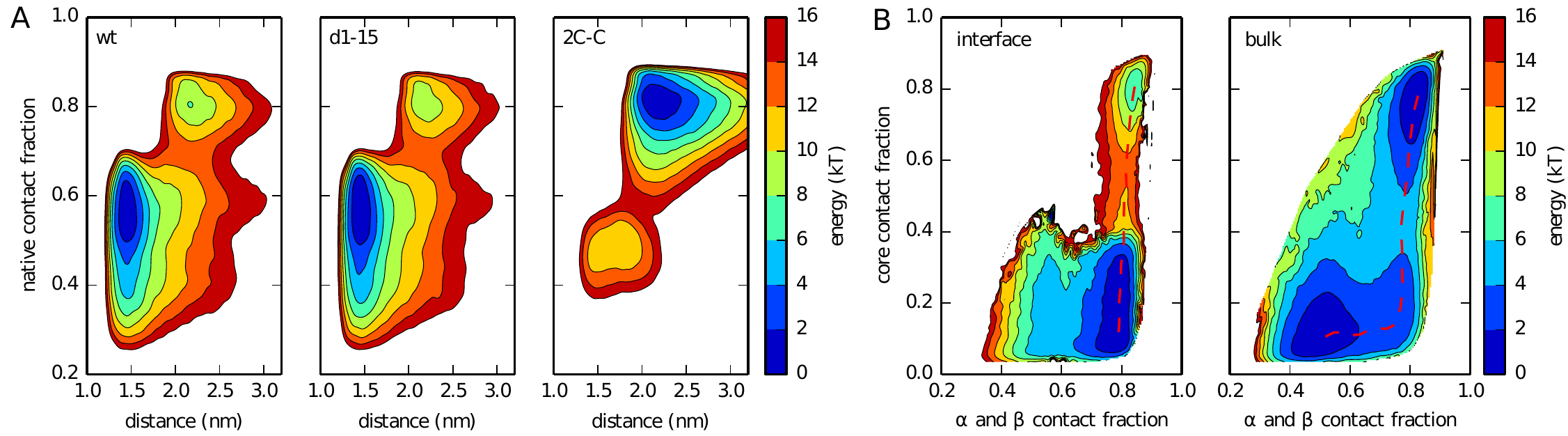}
\end{center}
\caption{A)Free energy landscape of the second step of adsorption for wt (left), d1-15 (center) and 2C-C (right) Rsn-2 as a function of the distance between the interface and the center of mass of Rsn-2$_{16-96}$ and the native contact fraction.
B) Comparison between the free energy landscape of wt Rsn-2 at the interface at ambient temperature (T=0.9T$_f$, left) and in bulk at the folding temperature (T=T$_f$, right). The dashed red line shows the minimum energy path between the two free-energy minima (corresponding to folded and unfolded states) as a function of the alpha and beta contact fraction and the core contact fraction .}
\label{fig_fes_step2}
\end{figure}

Figure~\ref{fig_fes_step2}A shows the three free energy landscapes of the wild-type protein and two mutant proteins unfolding at the interface as a function of the distance from the interface and the fraction of native contacts in the protein.
For all three proteins the landscapes display two free energy minima: one corresponding to the protein folded at the interface (state 1), and one corresponding to the protein unfolded and whose centre of mass is closer to the interface (state 2).
For wt and d1-15 Rsn-2 the increase in energy due to the loss of native contacts is widely compensated by the more favourable interactions with the interface, so that state 2 represents the global minimum of the system.
On the other hand, despite the existence of a local free energy minimum corresponding to the unfolded conformation, the introduction of the two additional disulphide bonds makes the unfolding of the core too unfavourable, and the system is trapped in its first adsorption state. This result is in agreement with the structure of the 2C-C mutant at the interface being identical to the solution state of wild-type Rsn-2 (Fig.~\ref{fig:RIMECD}).

It is interesting to compare the free energy landscape of the adsorbed protein with that of the protein in aqueous solution at the folding temperature, as a function of the fraction of secondary structure and hydrophobic core native contacts. 
Figure~\ref{fig_fes_step2}B shows that in both cases unfolding of Rsn-2 proceeds via the same minimum energy path, as computed using the nudged elastic band method~\cite{sheppard2008optimization}.
However, at the interface the unfolding is arrested in a state where the individual secondary structure elements are stable, whereas the hydrophobic contacts of the core are not.
The picture that emerges is that the structure of Rsn-2 at the interface is an arrested state on the unfolding pathway observed in the bulk.


\section{Discussion and conclusions}

In this work we have demonstrated, using both experiment and simulation, that Rsn-2 does undergo an unusual `clam-shell' like unhinging transition at an interface as first suggested by Mackenzie \emph{et al.} \cite{Mackenzie2009}. Furthermore, we have shown that Rsn-2 adsorption can be separated into two steps. First, we found that the unstructured, flexible hydrophobic N-terminal tail acts as a search mechanism, similar to the fly-casting mechanism suggested for intrinsically disordered proteins \cite{Shoemaker2000}, that can capture the interface without any structural reorganization. Therefore there is no energy barrier and initial adsorption is only diffusion controlled. Second, once at the interface, Rsn-2 undergoes a transition of its tertiary structure where the native contacts between the $\alpha$-helix and $\beta$-sheet in the native structure are replaced by equivalent hydrophobic contacts with the surface. Interestingly, however, the secondary structure is not lost partly due to the $\alpha$-helix and the $\beta$-sheet having a well-defined hydrophobic dipole.

While the experiments show that Rsn-2 does undergo this unusual transition, it cannot provide an answer to how and why, from an energetic perspective; only computational modelling can provide an answer. Our simulations show that the conformation Rsn-2 adopts upon adsorption to an interface is on the edge of stability. We have shown that this large scale transition follows the same energetic pathway as denaturation (Fig. 8B). When adsorbing, the interface acts as a perturbation that drives Rsn-2 down its unfolding pathway, but denaturation is arrested because the hydrophobic core contacts of the protein are energetically satisfied by the presence of the interface.

There do exist examples of other proteins that undergo a similar limited structural rearrangement at an interface. The apolipoproteins are the protein component of lipoprotein complexes and are known for their role in lipid transport processes \cite{Vance2008}. The apolipoproteins typically consist of four or five amphipathic $\alpha$-helices that form a helical bundle. It has been shown that these proteins can undergo a conformational switch where the protein undergoes a large scale conformational re-orientation whilst maintaining its overall secondary structure during the lipid binding process \cite{Narayanaswami2010, Wientzek1994}. In this case, the core hydrophobic helix-helix interactions are satisfied by helix-lipid interactions \cite{Leon2006}. Moreover, this binding process is reversible and is a key feature for its proper \emph{in vivo} functioning. In the case of apolipophorin-III produced by the insect \emph{Galleria mellonella}, such reversibility allows for a re-usable resource for lipoprotein transport. In mammals, the reversibility is thought to be potentially even more important as these apolipoproteins may play a role in a range of other cellular processes \cite{Narayanaswami2010}. Rsn-2 shares the same characteristics as the apolipoproteins when binding to an interface: a large conformational change with retention of secondary structure. As noted above, Rsn-2 surface adsorption appears to be reversible. This ability of Rsn-2 may allow the t\'{u}ngara frog to conserve resources (analogous to apolipophorin-III) or may indicate that Rsn-2 may play other roles \emph{in vivo}. Whether the conformational change associated with apolipoprotein binding to lipid membranes is also an arrested unfolding state is a question that our computational model would be ideally suited to explore. 

Rsn-2 shares close structural similarities to the family of cysteine proteinase inhibitors, the cystatins. These proteins are not known to have any surfactant characteristics while recombinant Rsn-2 has been shown not to possess any protease inhibition activity \cite{Fleming2009}. Cystatins have garnered interest due to their ability to form dimers via three-dimensional domain swapping. Such a mechanism may play a role in cystatin-related amyloidogenic diseases \cite{Staniforth2001}. During domain-swapped dimerization, the monomer undergoes a large-scale conformational change by an unhinging mechanism, exposing a large number of hydrophobic residues. This unhinged structure is exactly complemented by a similarly unhinged partner, thus forming homo-dimers with very similar structural characteristics to the monomer. The retention of secondary structural elements during domain swapping via unhinging was also shown to be a viable model in B1 domain of the immunoglobulin G binding protein (GB1) \cite{Malevanets2008}. While Rsn-2 and cystatin may not share similar functional roles \emph{in vivo}, the mechanism of three dimensional domain swapping found in the cystatins is remarkably similar to the mechanism of Rsn-2 surface adsorption. It poses an interesting question from an evolutionary perspective of how Rsn-2 may have transformed this domain swapping capability into its \emph{in vivo} role in the t\'{u}ngara frog. It also raises the question whether there are conditions or circumstances in which Rsn-2 could perform similar domain-swapping.    

It is thought Rsn-2 has no deleterious effects on biological membranes, making it an excellent candidate to act as a component in biocompatible soft materials. Indeed, it has been shown that Rsn-2, when incorporated with model cellular systems in both bulk solutions and foams, shows no disruptive activity to both cellular membranes and other proteins \cite{Choi2013}. Furthermore, Choi and co-workers demonstrated that foams formed from Rsn-2 are most stable at neutral pH making them ideally suited to act as architecture for biosynthesis applications and biocompatible materials. In fact, the biocompatibility of Rsn-2 foams has already been exploited as a platform in an artificial photosynthetic system \cite{Wendell2010}. 


Finally, the utility of the computational methods developed in this work could prove useful for understanding the conformational dynamics and thermodynamic driving processes of protein/interfacial interactions.
Due to its general nature, any interface may be modelled if the partitioning energies of the amino acids into the desired hydrophobic phase are known or can be estimated.
Therefore, expensive atomistic molecular dynamics simulations may be replaced by efficient coarse-grained simulations which still provide valuable information regarding the dynamical and thermodynamic processes of adsorption.
This method could then be applied to other known surfactant or lipid binding proteins.
In the current implementation, our model cannot be used to predict the formation of new secondary structure elements, e.g. the conversion of a disordered region into a $\beta$-sheet.
However, if the new structure at the interface is known from experiments, one could model the protein using a multiple-basin energy landscape~\cite{Okazaki2006}, and then study the coupling between the adsorption and the formation of the new secondary structure elements.

\section*{Acknowledgements}

We would like to thank Steven Vance for generating the mutant constructs. This work was funded by EPSRC grant EP/J007404.


\end{document}